\documentclass[twocolumn, prl, aps, floats]{revtex4}
\DeclareMathAlphabet{\mathpzc}{OT1}{pzc}{m}{it}

\usepackage{amsmath, amssymb, bbm, bm, color, epsf, epsfig, float, graphicx, ragged2e, relsize, setspace, tabularx, yfonts}
\def\lahigh{7ex}

\def\na{\overline{a}}
\def\nb{\overline{b}}
\def\nc{\overline{c}}

\def\ng{\overline{g}}

\def\Neta{\mbox{}\vspace{0pt}\hspace{0pt}		\includegraphics[height=\lahigh]{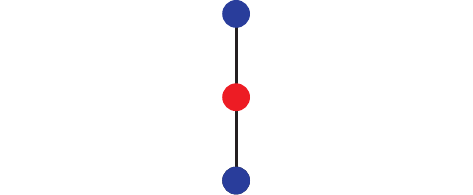}}
\def\Netaa{\mbox{}\vspace{0pt}\hspace{0pt}		\includegraphics[height=\lahigh]{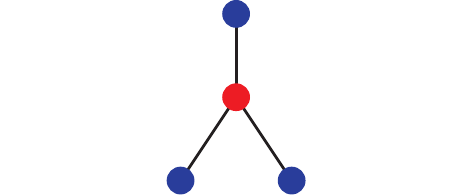}}
\def\Netaaa{\mbox{}\vspace{0pt}\hspace{0pt}		\includegraphics[height=\lahigh]{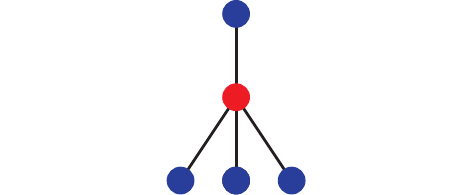}}
\def\Netab{\mbox{}\vspace{0pt}\hspace{0pt}		\includegraphics[height=\lahigh]{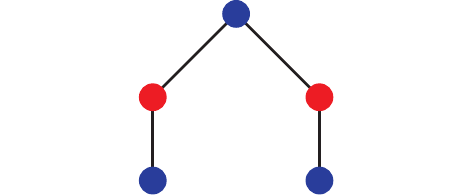}}
\def\Netabb{\mbox{}\vspace{0pt}\hspace{0pt}		\includegraphics[height=\lahigh]{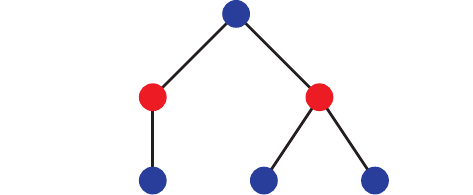}}
\def\Netaabb{\mbox{}\vspace{0pt}\hspace{0pt}		\includegraphics[height=\lahigh]{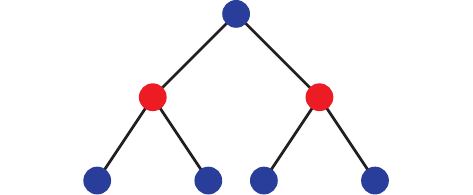}}
\def\Netabbb{\mbox{}\vspace{0pt}\hspace{0pt}		\includegraphics[height=\lahigh]{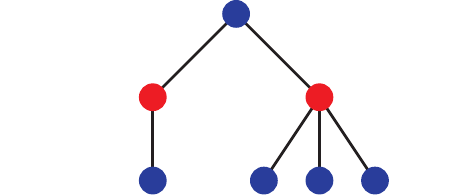}}
\def\Netaabbb{\mbox{}\vspace{0pt}\hspace{0pt}	\includegraphics[height=\lahigh]{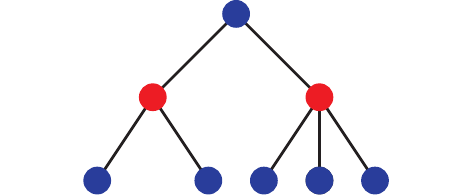}}
\def\Netaaabbb{\mbox{}\vspace{0pt}\hspace{0pt}	\includegraphics[height=\lahigh]{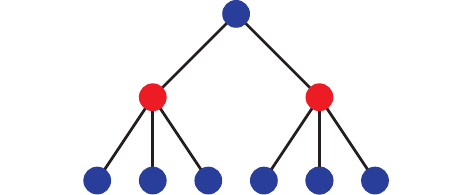}}
\def\Netabc{\mbox{}\vspace{0pt}\hspace{0pt}		\includegraphics[height=\lahigh]{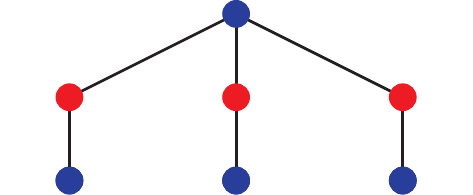}}
\def\Netabcc{\mbox{}\vspace{0pt}\hspace{0pt}		\includegraphics[height=\lahigh]{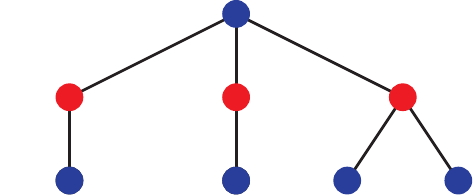}}
\def\Netabbcc{\mbox{}\vspace{0pt}\hspace{0pt}		\includegraphics[height=\lahigh]{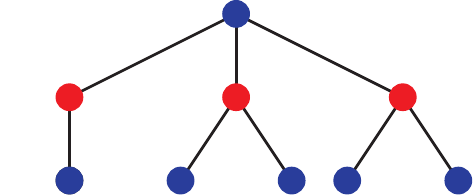}}
\def\Netabccc{\mbox{}\vspace{0pt}\hspace{0pt}		\includegraphics[height=\lahigh]{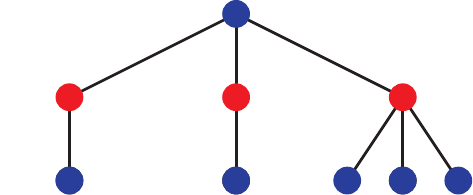}}
\def\Netabbccc{\mbox{}\vspace{0pt}\hspace{0pt}	\includegraphics[height=\lahigh]{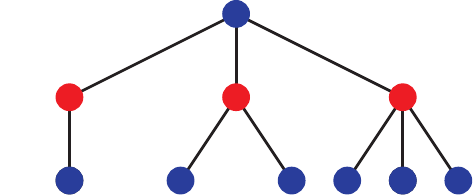}}
\def\Netabbbccc{\mbox{}\vspace{0pt}\hspace{0pt}	\includegraphics[height=\lahigh]{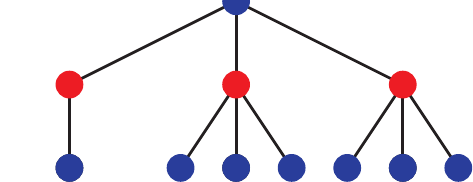}}
\def\Netaabbcc{\mbox{}\vspace{0pt}\hspace{0pt}	\includegraphics[height=\lahigh]{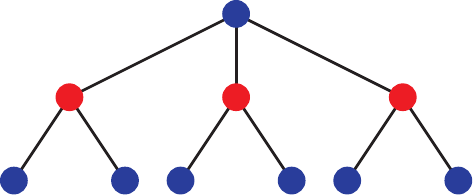}}
\def\Netaabbccc{\mbox{}\vspace{0pt}\hspace{0pt}	\includegraphics[height=\lahigh]{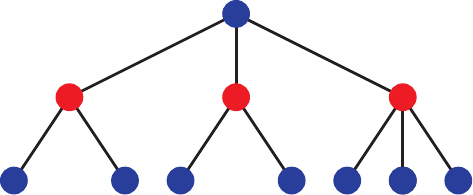}}
\def\Netaabbbccc{\mbox{}\vspace{0pt}\hspace{0pt}	\includegraphics[height=\lahigh]{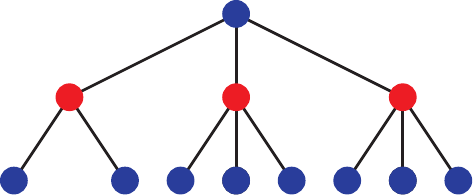}}
\def\Netaaabbbccc{\mbox{}\vspace{0pt}\hspace{0pt}	\includegraphics[height=\lahigh]{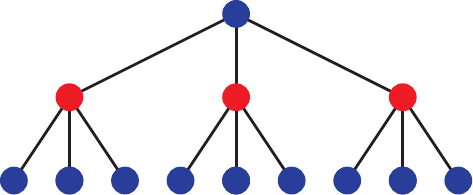}}

\def\ga{\mbox{}\vspace{0pt}\hspace{0pt}			\includegraphics[height=\lahigh]{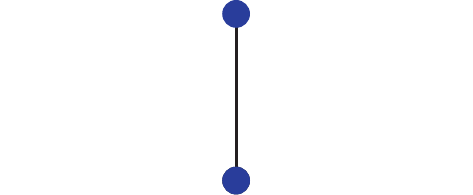}}
\def\gaa{\mbox{}\vspace{0pt}\hspace{0pt}			\includegraphics[height=\lahigh]{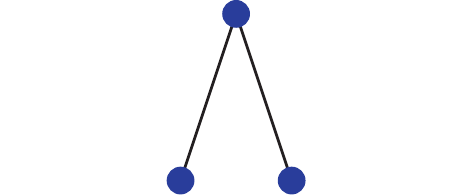}}
\def\gaaa{\mbox{}\vspace{0pt}\hspace{0pt}		\includegraphics[height=\lahigh]{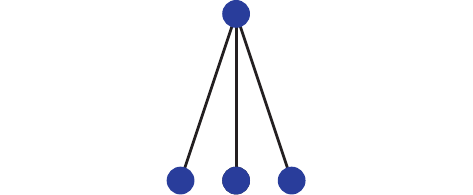}}
\def\gab{\mbox{}\vspace{0pt}\hspace{0pt}			\includegraphics[height=\lahigh]{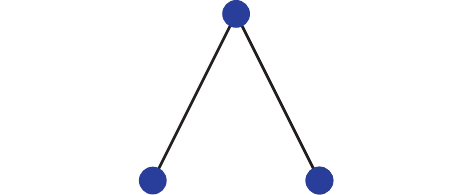}}
\def\gabb{\mbox{}\vspace{0pt}\hspace{0pt}		\includegraphics[height=\lahigh]{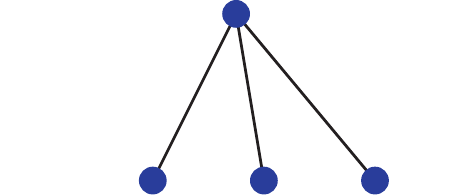}}
\def\gaabb{\mbox{}\vspace{0pt}\hspace{0pt}		\includegraphics[height=\lahigh]{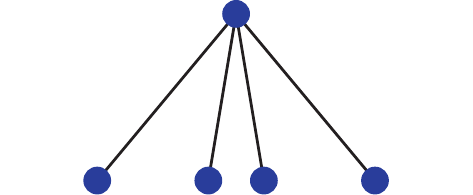}}
\def\gabbb{\mbox{}\vspace{0pt}\hspace{0pt}		\includegraphics[height=\lahigh]{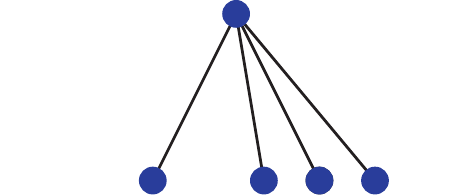}}
\def\gaabbb{\mbox{}\vspace{0pt}\hspace{0pt}		\includegraphics[height=\lahigh]{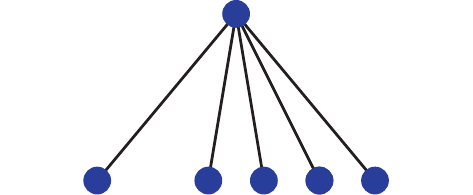}}
\def\gaaabbb{\mbox{}\vspace{0pt}\hspace{0pt}		\includegraphics[height=\lahigh]{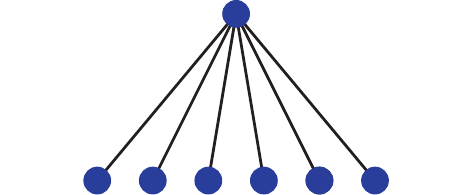}}
\def\gabc{\mbox{}\vspace{0pt}\hspace{0pt}		\includegraphics[height=\lahigh]{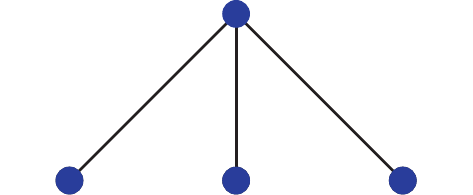}}
\def\gabcc{\mbox{}\vspace{0pt}\hspace{0pt}		\includegraphics[height=\lahigh]{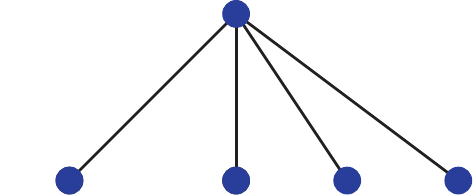}}
\def\gabbcc{\mbox{}\vspace{0pt}\hspace{0pt}		\includegraphics[height=\lahigh]{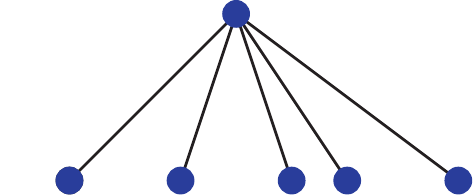}}
\def\gabccc{\mbox{}\vspace{0pt}\hspace{0pt}		\includegraphics[height=\lahigh]{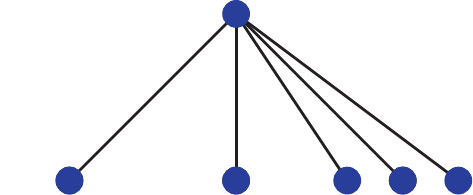}}
\def\gabbccc{\mbox{}\vspace{0pt}\hspace{0pt}		\includegraphics[height=\lahigh]{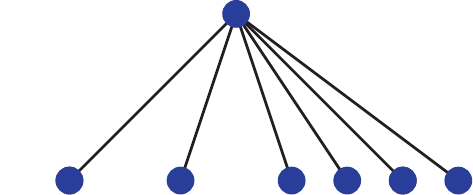}}
\def\gabbbccc{\mbox{}\vspace{0pt}\hspace{0pt}		\includegraphics[height=\lahigh]{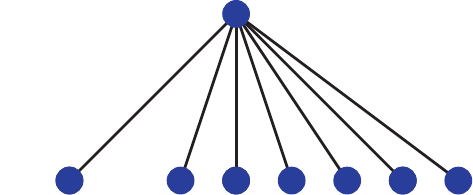}}
\def\gaabbcc{\mbox{}\vspace{0pt}\hspace{0pt}		\includegraphics[height=\lahigh]{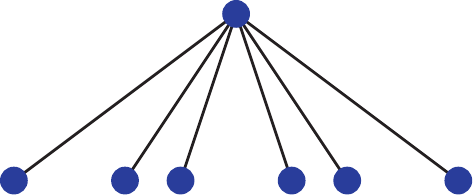}}
\def\gaabbccc{\mbox{}\vspace{0pt}\hspace{0pt}		\includegraphics[height=\lahigh]{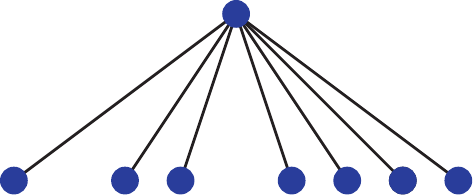}}
\def\gaabbbccc{\mbox{}\vspace{0pt}\hspace{0pt}	\includegraphics[height=\lahigh]{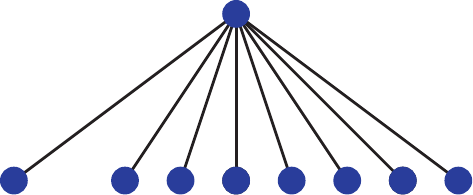}}
\def\gaaabbbccc{\mbox{}\vspace{0pt}\hspace{0pt}	\includegraphics[height=\lahigh]{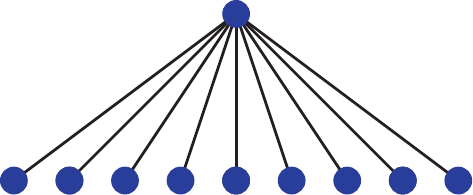}}

\begin{document}
	\begin{small}
		\title{\flushleft
			{\sf\LARGE \textbf{Regulatory motifs: structural and functional building blocks \\ of genetic computation}} \\
			\vspace{5pt}
			{\sf \textbf{T. M. A. Fink}} \\
			\vspace{3pt} 	
			{\sf \small \textbf{London Institute for Mathematical Sciences, Royal Institution, 21 Albermarle St, London W1S 4BS, UK}} \\
			\mbox{}				\vspace{-20pt} 
			\mbox{}
		}
		\maketitle
\noindent
\begin{justify}
{\sf\textbf{Developing and maintaining life requires a lot of computation. 
This is done by gene regulatory networks.
But we have little understanding of how this computation is organized.
I show that there is a direct correspondence between the structural and functional building blocks of regulatory networks,
which I call regulatory motifs.
I derive a simple bound on the range of function that these motifs can perform, in terms of the local network structure.
I prove that this range is a small fraction of all possible functions, which severely constrains global network behavior.
Part of this restriction is due to redundancy in the function that regulatory motifs can achieve---there are many ways to perform the same task. 
Regulatory motifs help us understanding how genetic computation is organized and what it can achieve.
}}
\end{justify}
\noindent
{\sf\textbf{\textcolor{red}{\large Introduction}}}
\\ \noindent {\sf\textbf{Genetic computation}} \\ 
Genetic computation is the computation done by gene regulatory networks.
It is responsible for 
the creation of body structures \cite{Gomez2008}, 
the determination of cell identity \cite{Pawlowski2017} and
resilience to fluctuations in the environment.
Scientists have investigated genetic computation for over 50 years \cite{Kauffman1969,Huang2005}.
Their work tends to be at two ends of a spectrum.
On the one hand, they have studied the average behavior of large networks.
This work, exemplified by Boolean networks \cite{Socolar2003,Samuelsson2003,Shmulevich2004}, 
considers the global dynamics of a large number interacting genes, typically with random connectivities and update rules.
On the other hand, scientists have studied the specific behavior of small networks.
This work, exemplified by the dynamics of network motifs \cite{Milo2002,Milo2004},
considers the specific function of small genetic circuits \cite{Mangan2003,Ahnert2016}, typically comprising just a handful of nodes.
\\ \indent 
Understanding how genetic computation is organized is one of the top mathematical challenges of our time \cite{Whipple2021}.
However, any comprehensive theory requires that we know how global behavior emerges from local function.
Biology makes extensive use of modularity: the repeated use of specific design elements that can be usefully combined \cite{Wagner2007}.
Modularity is likely to feature in genetic computation as well: the existence of functional subroutines that can be combined to create more advanced functionality.
We see this in modern software design, in which new software is rarely written from scratch.
Instead, it tends to combine existing subroutines, like how programs written in JavaScript combine a library of existing modular programs \cite{Decan2018}.
\\ \\ \noindent {\sf\textbf{Structure and function}} \\ 
In any system that performs computation, structure and function are different.
Structure dictates which parts of the system can interact, whereas function relates to the dynamics of the system as it evolves over time.
For example, 
in electronic computing, the structure is the circuit architecture;
in neural networks, it is the synapse connectivity;
in cellular automata, it is the neighbors that define the lattice;
in gene regulatory networks, it is the network connectivity of molecules that can bind together.
In contrast to structure, function takes place in dynamical space \cite{Wolfram2002a}. 
This is much larger than structure space, since the number of states grows exponentially with the size of the system.
\\ \indent
In general, at the most local level, structural and functional elements must correspond, 
because computation always takes place via a physical substrate---a physical dynamical system.
In electronic computing and neural networks, for example, these elements are the transistor and the neuron.
But as we progress to higher organizational length scales, this correspondence breaks down.
Instead, function gets distributed over different parts of the structure, 
like how a computer program is computed by different parts of a circuit.
\\ \indent
Any correspondence between structure and function in gene regulatory networks is valuable because it provides a window into how genetic computation is organized.
The essence of this paper is that there is a formal correspondence between the two that extends beyond individual molecular logics, 
and this tells us how genetic computation is built up and constrained. 
\\ \\ \noindent {\sf\textbf{Genes and transcription factors}} \\ 
One of the distinctive features of gene regulatory networks is their bipartite nature: genes and transcription factors talk to each other but not themselves \cite{Hannam2019,Hannam2019b,Fink2022}.
To see why, recall that a transcription factor is a protein or complex of proteins which are synthesized from expressed genes.
Thus a transcription factor depends on one or more genes.
A gene is a particular segment of DNA that codes for a protein, flanked by one or more binding sites for different transcription factors, which together promote or block the transcription of the gene.
Thus a gene depends on one or more transcription factors. 
In this way, the expression levels of genes are determined by those of other genes, but only indirectly---transcription factors act as middlemen 
(The same is true if we swap genes and transcription factors, but I stick to the gene-centric perspective for simplicity.)
\\ \indent
In previous work \cite{Fink2022}, I showed that a consequence of this bipartite nature is that the logical dependencies that one gene can have on other genes---what I call biological logics---are restricted.
In other words, many of the possible logical dependencies are forbidden.
I was able to enumerate these biological logics, some of which are given in \cite{Fink2022}, but I was unable to derive a simple expression for their number.
By casting the problem in terms of structural and functional building blocks,
in this work I derive a deeper understanding of the restriction on biological logics, and how they combine to perform more advanced computation.
\def\vs{\vspace{0.02in}}
\def\lahigh{4.85ex}
\begin{figure*}[t!]
\includegraphics[width=0.91\columnwidth]{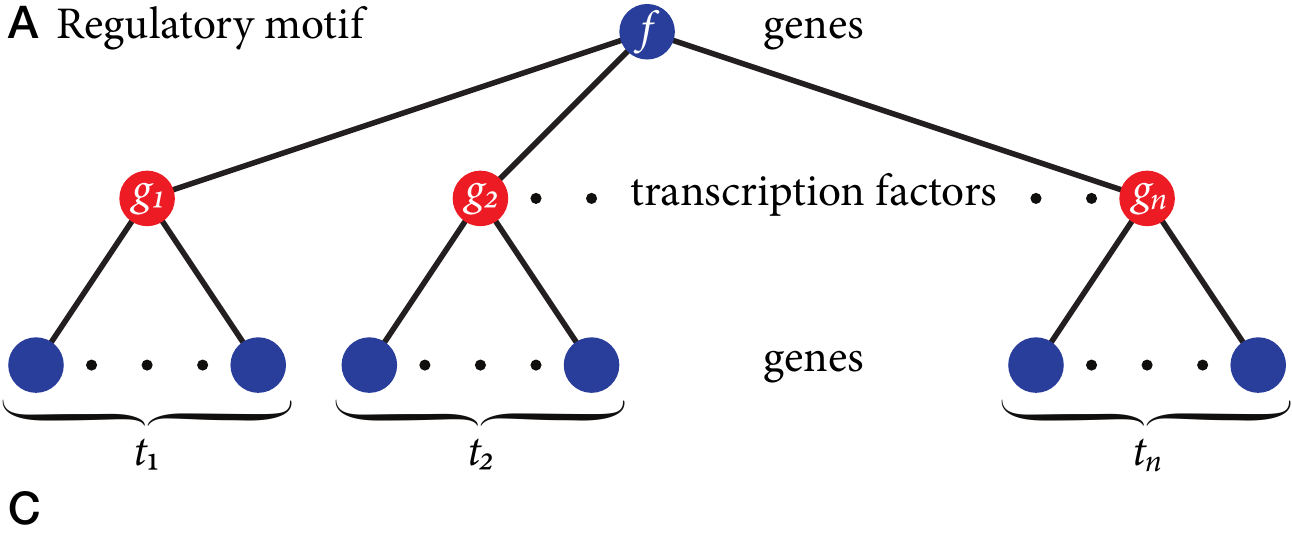} 
\hspace{0.7in}
\includegraphics[width=0.91\columnwidth]{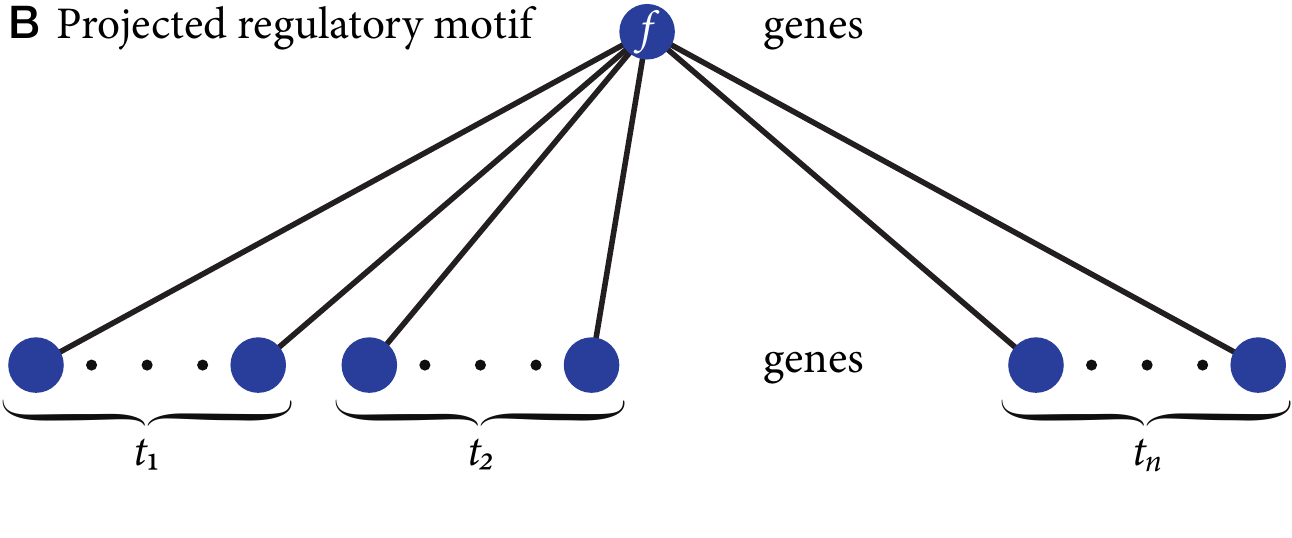}
\setlength{\tabcolsep}{0pt}
\begin{tabularx}{\textwidth}{@{\extracolsep{\fill}}cccccccc}
\emph{Short-}	& \emph{Regulatory}		& \emph{Lower}	& \emph{Biological}	& \emph{Uppper}			&& \emph{Projected}		& \emph{All}					\\	
\emph{hand}	& \emph{motifs}		& \emph{bound}	& \emph{logics}		& \emph{bound}			&& \emph{regulatory}	& \emph{logics}					\\	
			&					& $c_{\rm lower}$	& $c$			& $c_{\rm upper}$			&& \emph{motifs}		& $\ell$						\\	
$\{1\}$		& \Neta				& 2				&  4				& 4						&& \ga  	 			& 4        					\vs	\\
$\{2\}$		& \Netaa				& 14				& 16   			& 16						&& \gaa				& 16        					\vs	\\
$\{3\}$ 		& \Netaaa				& 254			& 256   			& 256					&& \gaaa				& 256        				\vs	\\
$\{1,1\}$ 		& \Netab				& 10				& 16    			& 40						&& \gab				& $16$   		   			\vs	\\
$\{1,2\}$		& \Netabb				& 70				& 88     			& 160					&& \gabb				& $256$   					\vs	\\
$\{2,2\}$		&\Netaabb			& 490			& 520    			& 640					&& \gaabb				& $65,536$   	 			\vs	\\
$\{1,3\}$		& \Netabbb			& 1,270			& 1,528   		  	& 2,560					&& \gabbb				& $65,536$                       		\vs	\\
$\{2,3\}$		& \Netaabbb			& 8,890			& 9,160 			& 10,240					&& \gaabbb			& $4.3 \times 10^9$ 			\vs	\\
$\{3,3\}$		& \Netaaabbb			& 161,290			& 161,800    		& 163,840			  		&& \gaaabbb			& $1.8 \times 10^{19}$ 		\vs	\\
$\{1,1,1\}$		& \Netabc				& 218			& 256 			& 1,744		 			&& \gabc				& 256         	  			\vs	\\
$\{1,1,2\}$		& \Netabcc			& 1,526			& 1,696   			& 6,976					&& \gabcc				& 65,536      			  	\vs	\\
$\{1,2,2\}$		& \Netabbcc			& 10,682			& 11,344   		& 27,904					&& \gabbcc			& $4.3 \times 10^9$   	 	\vs	\\
$\{1,1,3\}$ 	& \Netabccc			& 27,686			& 30,496  			& 111,616					&& \gabccc			& $4.3 \times 10^9$       		\vs	\\
$\{2,2,2\}$		& \Netaabbcc			& 74,774			& 76,288 			& 111,616					&& \gaabbcc 			& $1.8 \times 10^{19}$  		\vs	\\
$\{1,2,3\}$		& \Netabbccc			& 193,802			&  204,304 		& 446,464					&& \gabbccc			& $1.8 \times 10^{19}$		\vs	\\
$\{2,2,3\}$ 	& \Netaabbccc			& 1,356,614		& 1,375,168		& 1,785,856				&& \gaabbccc			& $3.4 \times 10^{38}$ 	 	\vs 	\\
$\{1,3,3\}$ 	& \Netabbbccc			& 3,516,122		& 3,680,464 		& 7,143,424				&& \gabbbccc			& $1.8 \times 10^{19}$       	\vs	\\
$\{2,3,3\}$ 	& \Netaabbbccc		& 24,612,854		& 24,792,448 		& 28,573,696				&& \gaabbbccc			& $1.2 \times 10^{77}$  	 	\vs	\\
$\{3,3,3\}$ 	& \Netaaabbbccc    		& 446,547,494		& 447,032,128		& 457,179,136				&& \gaaabbbccc		& $1.3 \times 10^{154}$			
\end{tabularx}%
\caption{
\textbf{Regulatory motifs and their projections, and the number of logics they can support.}
\textbf{A}
In a regulatory motif, a gene (blue) depends on $n$ transcription factors (red), each of which depends on $t_1,\ldots,t_n$ genes (blue). 
As a shorthand, we write $\{t_1,\ldots,t_n\}$, which counts the number of genes in the $n$ branches of the tree.
\textbf{B}
In a projected motif, a gene depends on $t_1+\ldots+t_n$ genes; there are no transcription factor middlemen.
\textbf{C}
For the 19 simplest regulatory motifs (left), I show
the number of biological logics $c$, and the lower and upper bounds on $c$.
For their projections (right), I show the number of all possible logics $\ell$.
This tends to be much larger than the number of biological logics $c$. 
}
\label{MainTable}
\end{figure*}
\\ \\ \noindent {\sf\textbf{Logics}} \\ 
I take genes and transcription factors to be either expressed or not expressed.
This means that the expression of a gene is a Boolean function of the transcription factors that regulate it, 
and the expression of a transcription factor is a Boolean function of the genes that code for its proteins.
A Boolean function is just a logic gate---a lookup table for the state of the output given the states of the inputs.
I refer to a Boolean function and a logic interchangeably, with a bias towards logic, which is simpler. 
(Debate about discrete versus continuous expression neglects a deeper concern over whether sophisticated continuous computation is even possible \cite{Wolfram2002b}.)
\\ \indent
There are $2^{2^n}$ logics of $n$ inputs.
For example, for $n=2$, these are
true,  false, 
$a$, $b$, $\na$, $\nb$
$ab$, $a\nb$, $\na b$, $\na\nb$,
$a+b$, $a+\nb$, $\na+b$, $\na+\nb$,
$ab+\na\nb$ and $a\nb+\na b$. 
In this notation, $\na$ means {\sc not} a, $ab$ means $a$ {\sc and} $b$, and $a+b$ means $a$ {\sc or} $b$.
Notice that two of these 16 logics depend on no inputs,
four depend on one input, 
and 10 depend on two inputs.
As we shall see, an important quantity is the number of logics of $n$ inputs which depend on all $n$ inputs, which we call $s(n)$, shown in eq. (\ref{Sn}).
The first few $s(n)$ are 2, 2, 10, 218, 64594 (OEIS A000371 \cite{Sloane}), starting at $n=0$. 
\\ \\ \noindent {\sf\textbf{In this paper}} \\ 
In this paper I do four things, which correspond to the four parts of the Results.
First, I show that gene regulatory networks can be uniquely broken into structural building blocks (Fig. \ref{MainTable}C left), 
and that these correspond to the functional building blocks of the network.
Just as the global network structure is a combination of these regulatory motifs,
the global network function is a combination of the biological logics that these motifs can carry out.
Second, I study the number of different biological logics that these regulatory motifs can perform.
By bounding it from below and above, I provide a simple estimate of this number in terms of the local network structure.
Third, I prove that this number is a tiny fraction of the number of all possible logics of the same number of inputs. 
This puts severe constraints on the global behavior of regulatory networks.
Fourth, I prove that part of the restriction on biological logics is due to their redundancy: 
different assignments of update rules to the nodes in a regulatory motif can produce the same biological logic.
I calculate the average redundancy in terms of the local network structure.
I conclude in the Discussion with implications of these results on the organization of genetic computation.
\\ \\ \noindent {\sf\textbf{\textcolor{red}{\large Results}}} \\
\noindent {\sf\textbf{Regulatory motifs: structural and functional building blocks}}\\ 
I start by showing there is a direct correspondence between the structural and functional building blocks of regulatory networks. 
Let's first look at the structure.
The primitive structural building blocks are the connectivities that a gene has with other genes via transcription factor middlemen.
For a gene that depends on $n$ transcription factors, each of which depends on $t_1,\ldots,t_n$ genes (Fig. \ref{MainTable}A),  
we use as a shorthand $\{t_1,\ldots,t_n\}$. 
For example, $\{2,2\}$ denotes a gene which depends on two transcription factors, each of which depends on two genes.
We call these pieces regulatory motifs, the 19 simplest of which are shown in Fig. \ref{MainTable}C left.
Any bipartite network can be broken into such pieces in a unique way.
The recipe for doing so is to pick a gene (blue node), find all its second-nearest neighbors, then break each of these outer genes in half.
(I could have drawn the building blocks in Fig. \ref{MainTable} with the genes as half-nodes, but I kept them whole for convenience.)
\\ \indent
Now let's consider the functionality of these regulatory motifs.
The top gene in Fig. \ref{MainTable}A is a function of its transcription factors, which in turn are functions of their genes.
Ultimately, the state of the top gene is set by the state of the bottom genes.
We can work out this dependence by composing the logics. 
Let $x_{i,j}$ be the state of the $j$th gene in the $i$th branch in Fig. \ref{MainTable}A.
Then we can compose the logics to get a unique new logic:
\begin{equation}
\begin{aligned}
& h(x_{1,1},\ldots,x_{1,t_1}; \ldots; x_{n,1},\ldots,x_{n,t_n}) = \\
& f\big(g_1(x_{1,1},\ldots,x_{1,t_1}),\ldots,g_n(x_{n,1},\ldots,x_{n,t_n})\big).
\end{aligned}
\label{composed}
\end{equation}
\indent
For example, consider the regulatory motif $\{1,2\}$.
With the $x_{i,j}$ denoted $a,b,c$ for convenience, $h(a,b,c) = f(g_1(a),g_2(b,c))$.
Setting $f = g_1$ {\sc and} $g_2$, $g_1 = a$, and $g_2 = b$ {\sc or} $c$, 
the composed logic is $h = (a$ {\sc and} $b$) {\sc or} ($a$ {\sc and} $c$).
Different choices of the logics for $f$ and for $g_1$ and $g_2$ can give other composed logics.
\\ \\ \noindent {\sf\textbf{\textcolor{black}{Number of functions that a regulatory motif can perform}}}
\\ 
The number of biological logics $c$ is the number of logical dependencies that one gene can have on other genes, 
that is, the number of different forms that $h$ can take.
My main mathematical result is that $c$ is bounded by
\begin{eqnarray}
	c(t_1,\ldots,t_n) &\geq& s(n) \prod_{i=1}^n \!\left(2^{2^{t_i}}\!\!/2 - 1\right) = c_{\rm lower} 
	\label{lowerbound} \\
	c(t_1,\ldots,t_n) &\leq& s(n) \prod_{i=1}^n 2^{2^{t_i}}\!\!/2  =  c_{\rm upper},
	\label{upperbound}
\end{eqnarray}
where $s(n)$ is the number of logics of $n$ inputs which depend on all $n$ inputs,
\begin{equation}
    s(n) = \sum_{i=0}^n (-1)^{n-i} \binom{n}{i}2^{2^i}.
    \label{Sn}
\end{equation}	
These bounds are derived in the Methods.
For example, for the regulatory motif $\{2,3\}$, 
$c_{\rm upper} = s(2) \cdot 2^{2^2}\!\!/2 \cdot 2^{2^3}\!\!/2 =$ 10,240, which is not much higher than the true value, 9,160.
Values of $c$ and its bounds are given in Fig. \ref{MainTable}C and plotted as error bars in Fig. \ref{FigCompare}.
\begin{figure*}[t!]
	\centering
	\includegraphics[width=1\textwidth]{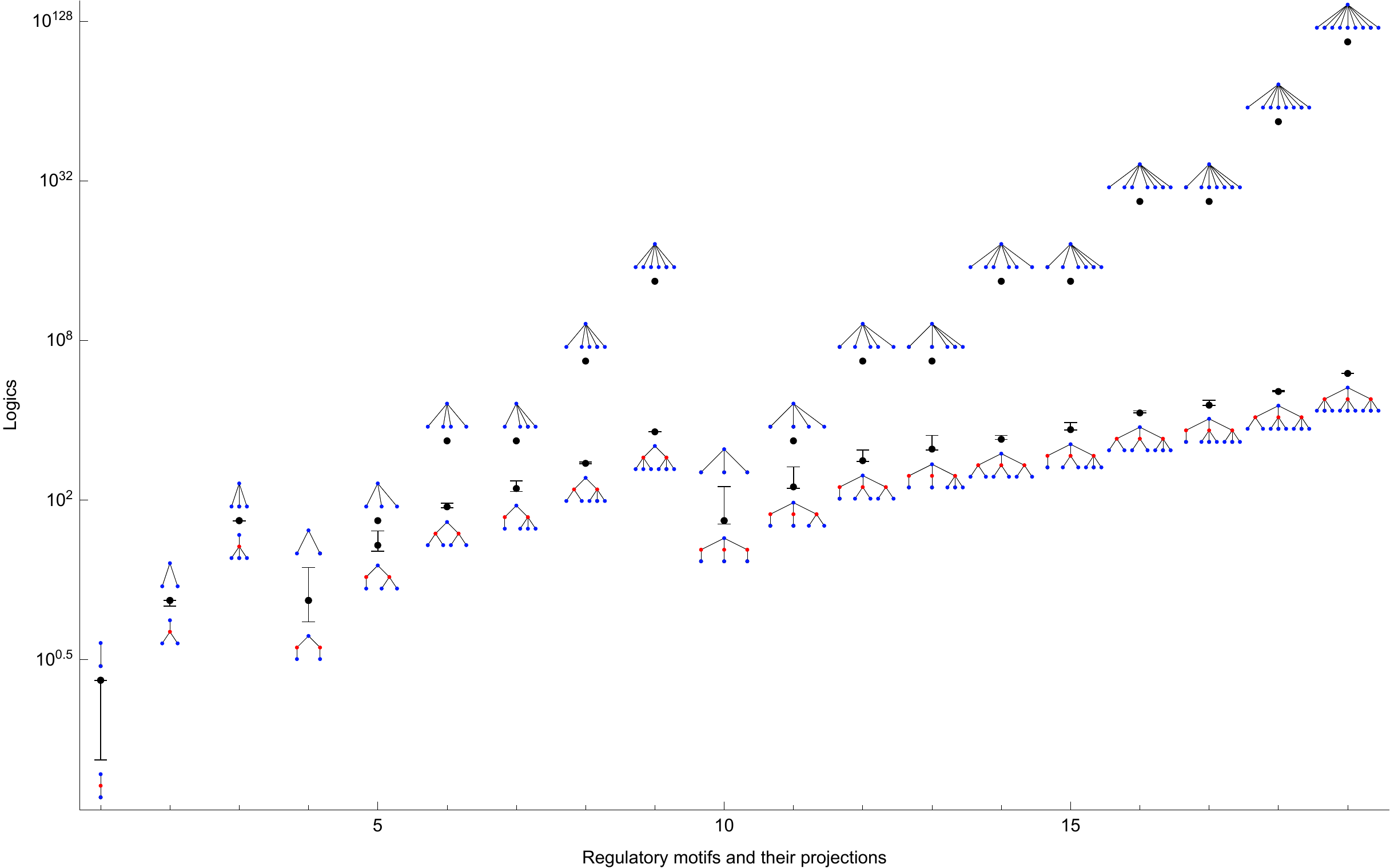}
	\caption{
	\textbf{Plot of the number of logics for regulatory motifs and their projections.}
	Each row in Fig. \ref{MainTable}C is plotted along the $y$-axis, in the same order that the rows appear in Fig. \ref{MainTable}C.
	The bounds for the number of biological logics $c$ are shown as points with error bars.
	The number of all possible logics $\ell$ are shown as points.
    }
    \label{FigCompare}
\end{figure*}
\\ \indent
The upper bound on the number of biological logics is also a good approximation to it.
We know this because the ratio of the lower and upper bounds approaches one as the $t_i$ increase.
Dividing eqs. (\ref{lowerbound}) and (\ref{upperbound}) by (\ref{upperbound}), the ratio of the bounds is 
\begin{equation*}
\frac{c_{\rm lower}}{c_{\rm upper}} = \prod_{i=1}^n \left( 1 - 2/2^{2^{t_i}}\right).
\end{equation*}
\indent
For example, for the regulatory motif $\{3,3\}$, the ratio is $(1 - 2/2^{2^3})^2$ = 0.984.
So each bound is within 1.6\% of $c$ itself.
\\ \\ \noindent {\sf\textbf{Function is restricted}} \\
In the absence of the transcription factor middlemen, we can simplify the regulatory motifs by projecting them: 
drawing an edge between genes connected to the same transcription factor.
The projections are shown in Fig. \ref{MainTable}C right.
\\ \indent
The top gene in Fig. \ref{MainTable}B depends on $t_1+\ldots+t_n$ other genes.
Since the number of logics of $n$ inputs is $2^{2^n}\!$,
the number $\ell$ of all logical dependencies that a gene can have in the absence of the transcription factors is 
\begin{equation*}
\ell = 2^{2^{t_1+\ldots+t_n}}.
\end{equation*}
\indent 
The range of functionality $c$ for the regulatory motifs is much smaller than the range $\ell$ for their projections 
(see Fig. \ref{MainTable}C and Fig. \ref{FigCompare}).
Our upper bound on $c$ lets us quantify this difference:
\begin{equation}
	\frac{c}{\ell} \leq \frac{c_{\rm upper}}{\ell} =
	\frac{s(n)}{2^n}
\frac{2^{2^{t_1}}\ldots 2^{2^{t_n}}}{2^{2^{t_1+\ldots+t_n}}}.
\label{RestrictedEq}
\end{equation}
Taking the logarithm gives a better sense of this ratio:
\begin{equation*}
	\log_2\left(\frac{c}{\ell}\right) \leq 2^n - n + \sum_{i=1}^n 2^{t_i} - \prod_{i=1}^n 2^{t_i},
\end{equation*}
where we make use of $s(n) \leq 2^{2^n}$\! (see Methods).
\\ \indent
For example, for the regulatory motif $\{2,3\}$, the number of biological logics $c$ is 9,160 and the number of possible logics $\ell$ is $2^{32}$.
Then $\log_2(c/\ell) = -18.8$, which is less than the bound of $2^2 - 2 + 2^2 +2^3 - 2^2 2^3 = -18$.
\\ \\ \noindent {\sf\textbf{\textcolor{black}{Function is redundant}}} \\
We know that the number of logical dependencies that one gene can have on other genes cannot be greater than the number of assignments of logics to the gene and transcription factors,
that is, the number of choices of $f$ and $g_1, \ldots, g_n$ in Fig. \ref{MainTable}A.
Since the number of logics of $n$ inputs is $2^{2^n}$,
the number of assignments is 
\begin{equation*}
v= 2^{2^n} 2^{2^{t_1}}\!\ldots 2^{2^{t_n}}.
\end{equation*}
Let's compare $v$ to the number of biological logics $c$: 
\begin{equation}
	\frac{v}{c} \geq \frac{v}{c_{\rm upper}} = 2^n \frac{2^{2^n}}{s(n)} \geq 2^n,
	\label{redundant}
\end{equation}
where we make use of $s(n) \leq 2^{2^n}$. 
So the average redundancy is at least $2^n$---though it can be considerably larger than this.
Values of the average redundancy are shown for the 19 simplest regulatory motifs in Fig. \ref{Redundancy}.
One instance of redundancy is that there is another way to compose the logic $h = (a$ {\sc and} $b$) {\sc or} ($a$ {\sc and} $c$) mentioned above.
It also results from setting $f = \ng_1$ {\sc and} $\ng_2$, $g_1 = \na$, and $g_2 = \nb$ {\sc and} $\nc$.
\\ \indent
For example, consider the regulatory motif $\{2,2\}$.
The number of assignments $v = 2^{2^2} 2^{2^2} 2^{2^2}$ = 4,096 and the number of biological logics $c$ is 520.
So the average redundancy is 7.9, which is indeed greater than $v/c_{\rm upper}$, which is $2^2 2^{2^2}/10 =$ 6.4
\def\vs{\vspace{0.01in}}
\def\lahigh{3.8ex}
\begin{figure}[t!]
\setlength{\tabcolsep}{0pt}
\begin{tabularx}{\columnwidth}{@{\extracolsep{\fill}}ccccc}
\emph{Short-}	& \emph{Regulatory}		& \emph{Logic}			& \emph{Biological}		& \emph{Average}		\\	
\emph{hand}	& \emph{motifs}		& \emph{assignments}	& \emph{logics}			& \emph{redundancy}	\\	
			& 					& $v$				& $c$				& $v/c$ 				\\	
$\{1\}$		& \Neta				& $2^4$				& 4					& 4			\vs		\\
$\{2\}$		& \Netaa				& $2^6$				& 16					& 4			\vs		\\
$\{3\}$		& \Netaaa				& $2^{10}$			& 256				& 4			\vs		\\
$\{1, 1\}$		& \Netab				& $2^8$				& 16					& 16			\vs		\\
$\{1, 2\}$		& \Netabb				& $2^{10}$			& 88					& 11.6		\vs		\\
$\{2, 2\}$		& \Netaabb			& $2^{12}$			& 520				& 7.9			\vs		\\
$\{1, 3\}$		& \Netabbb			& $2^{14}$			& 1,528				& 10.7		\vs		\\
$\{2, 3\}$		& \Netaabbb			& $2^{16}$			& 9,160				& 7.2			\vs		\\
$\{3, 3\}$		& \Netaaabbb			& $2^{20}$			& 161,800				& 6.5			\vs		\\
$\{1, 1, 1\}$	& \Netabc				& $2^{14}$			& 256				& 64			\vs		\\
$\{1, 1, 2\}$	& \Netabcc			& $2^{16}$			& 1,696				& 38.6		\vs		\\
$\{1, 2, 2\}$	& \Netabbcc			& $2^{18}$			& 11,344				& 23.1		\vs		\\
$\{1, 1, 3\}$	& \Netabccc			& $2^{20}$			& 30,496				& 34.4		\vs		\\
$\{2, 2, 2\}$	& \Netaabbcc			& $2^{20}$			& 76,288				& 13.7		\vs		\\
$\{1, 2, 3\}$	& \Netabbccc			& $2^{22}$			& 204,304				& 20.5		\vs		\\
$\{2, 2, 3\}$	& \Netaabbccc			& $2^{24}$			& 1,375,168			& 12.2		\vs		\\
$\{1, 3, 3\}$	& \Netabbbccc			& $2^{26}$			& 3,680,464			& 18.2		\vs		\\
$\{2, 3, 3\}$	& \Netaabbbccc		& $2^{28}$			& 24,792,448			& 10.8		\vs		\\
$\{3, 3, 3\}$	& \Netaaabbbccc		& $2^{32}$			& 447,032,128			& 9.6	
\end{tabularx}%
\caption{
\textbf{The average redundancy for regulatory motifs.}
For each of the 19 simplest regulatory motifs, I show:
the number of ways $v$ of assigning logics to the gene and transcription factors ($f$ and the $g_i$ in Fig. \ref{MainTable}A);
the number of distinct logics $c$ that these compose to (what I call biological logics); and 
their ratio, which is the average redundancy of this many-to-one map.
The average redundancy is at least $2^n$, where $n$ is the number of branches, but it can be considerably higher than this.
}
\label{Redundancy}
\end{figure}
\\ \\ \noindent {\sf\textbf{\textcolor{red}{\large Discussion}}} \\
\noindent {\sf\textbf{Correspondence of structure and function}} \\
In general, for any network that performs computation---genetic or otherwise---functional subroutines do not correspond to network substructures.
Rather, function tends to be distributed over different parts of the structure. 
However, for gene regulatory networks, structure and function do correspond at the length scale of regulatory motifs.
The functional subroutines---what I call biological logics---are precisely those that run on these network substructures.
Just as global network structure is built from the regulatory motifs in Fig. \ref{MainTable}C left,
global network function is built from the logics that run on them.
\\ \indent
The reason for this correspondence is because the network is bipartite: genes and transcription factors talk to each other but not themselves \cite{Hannam2019,Hannam2019b,Fink2022}.
In the building blocks in Fig. \ref{MainTable}C, all of the downstream transcription factors (red nodes) are surrounded by genes (blue nodes). 
This allows us to effectively integrate out the transcription factor logics, generating a restricted range of logical dependencies that the top gene can have on the bottom genes.
Incidentally, bipartite networks are not the only kind of networks for which structure and function correspond.
Tripartite networks and more complex network architectures also have this feature.
\\ \\ \noindent {\sf\textbf{Function is restricted}} \\
The extent to which biological logics are restricted is bounded by eq. (\ref{RestrictedEq}).
For $n=1$, or for all of the $t_i=1$, there is no restriction;
the number of biological and possible logics are the same, and $c/\ell = 1$.
But the restriction quickly becomes severe for other regulatory motifs.
The highest values that $c/\ell$ can take after 1 are 11/32 for the motif $\{1,2\}$ and 0.026 for $\{1,1,2\}$.
The ratio drops off for other regulatory motifs, reaching $10^{-14}$ for $\{2,2,2\}$ and $10^{-145}$ for $\{3,3,3\}$.
When all of the $t_i = t$, the logarithm of $c/\ell$ is less than $2^n + n(2^t-1) - 2^{nt}$. 
\\ \indent
For most of the regulatory motifs in Fig. \ref{MainTable}C, the restriction on function $c/\ell$ is astronomical.
This puts severe constraints on the global function of gene regulatory networks,
because the global function is a composition of the local function performed by each of the regulatory motifs.
As the number of motifs that are combined increases, the restriction on global behavior grows exponentially.
\\ \\ \noindent {\sf\textbf{Function is redundant}} \\
We know from eq. (\ref{redundant}) that biological logics are on average redundant.
If we think of the assignment of logics to the genes and transcription factors as the genotype (instructions),
and the composed gene-gene logic as the phenotype (behavior),
then the ratio of the number of genotypes to the number of phenotypes is at least $2^n 2^{2^n}\!/s(n)$, which is itself at least $2^n$.
The mapping of many genotypes to fewer phenotypes is highly prevalent in biological systems, 
ranging from RNA secondary structure to protein folding to biological clocks.
This gives rise to neutral networks and can confer robustness to errors \cite{Wagner2008}.
\\ \indent
Intriguingly, computational evidence suggests that some biological logics are much more redundant than others.
These phenotypes are more likely to be observed simply because there are more ways to design them.
For example, for the regulatory motif $\{2,3\}$, $2^{16} =$ 65,536 assignments map to 9,160 biological logics.
But of the $2^{16}$ assignments, 
18\% map to true and false, 
11\% map to 14 simple logics such as $a$, $b$, and $a$ {\sc or} $b$,
17\% map to 254 more complex logics, such as $c$ {\sc or} $d$ {\sc or} $e$,
and the rest map to 8,890 more complex logics still.
If this trend holds for regulatory motifs in general, it would imply that biological logics are not only restricted, but also tend to be simple.
This is an open theoretical question with important consequences, and I hope that others will try to answer it.
\\ \indent
More generally, the composition of logics is an archetypal system for understanding input-output maps.
Across a broad range of systems in nature and mathematics, input-output maps tend to be highly biased towards simple outputs \cite{Dingle2018,Johnston2022}.
A theory for the distribution of redundancies mentioned above could help provide an explanation of this empirical trend.
\\ \\ \noindent {\sf\textbf{\textcolor{red}{\large Methods}}} \\
{\sf\textbf{\textcolor{black}{Logics that depend on all of their inputs}}} \\
By the principle of inclusion and exclusion, the number of logics of $n$ inputs which depend on all $n$ inputs is 
\begin{equation}
    s(n) = \sum_{i=0}^n (-1)^{n-i} \binom{n}{i}2^{2^i},
    \label{SLogics}
\end{equation}	
which is the inverse binomial transform of $2^{2^n}\!$ (OEIS A000371 \cite{Sloane}).
It is bounded from below and above by
\begin{equation}
	2^{2^n} - n \, 2^{2^{n-1}} \leq s(n) \leq 2^{2^n}.
	\label{SBounds}
\end{equation}
The right side follows from the definition of $s(n)$.
The left side can be deduced by showing that the magnitude of the $(i-1)$th term in (\ref{SLogics}) is less than half that of the $i$th term, that is,
\begin{equation*}
	\binom{n}{i-1}2^{2^{i-1}} \leq \frac{1}{2} \binom{n}{i}2^{2^i}.
	\label{FD}
\end{equation*}
This implies that 
$2 i \leq (n-i+1) 2^{2^{i-1}}$.
Since $n-i+1$ is at least 1, we need only that $2 i \leq 2^{2^{i-1}}$.
It is indeed for all $i \geq 1$, establishing the lower bound in (\ref{SBounds}).
It implies that $s(n)$ rapidly approaches $2^{2^n}$.
\\ \\ \noindent {\sf\textbf{\textcolor{black}{Proof of upper bound on biological logics}}} \\
The upper bound in eq. (\ref{upperbound}) is not at all obvious.
It is also delicate, in that it is an equality for $n=1$. 
My first proof of the bound, not given here, was cumbersome.
But, like a climber who on reaching the summit sees a superior path of ascent, 
as soon as I proved it I found a much simpler proof the next day.
I thank my colleague Yang-Hui He for prompting one of the steps.
\\ \indent
Here is the simpler proof. 
My starting point is a result from a recent paper \cite{Fink2022}, in which I derived an exact---but difficult to apply---expression for the number of biological logics.
It is
\begin{equation}
    c(t_1,\ldots,t_n)  =  \sum_{m=0}^n s(m)	\!\! \sum_{\sigma_1\ldots \sigma_m}  \alpha_{\sigma_1} \ldots \alpha_{\sigma_n},  
    \label{cOriginal}   
\end{equation}
where 
\begin{equation*}
    \alpha_i =  2^{2^i}\!\!/2 - 1,
\end{equation*}
and the second sum adds up the product of all $m$-tuples of the $\alpha_i$. 
A few examples illustrate what this means:
\begin{eqnarray*}
    c(i)        &=& 2 + 2 \, \alpha_i, 												\label{HH} 	\\
    c(i,j)     	&=& 2 + 2(\alpha_i + \alpha_j)              + 10 \, \alpha_i \alpha_j, 				\label{HI} 		\\
    c(i,j,k)   	&=& 2 + 2(\alpha_i + \alpha_j + \alpha_k)  								\label{HJ} 	\\
    		&+& 10 \big(\alpha_i \alpha_j + \alpha_j \alpha_k + \alpha_i \alpha_k \big) + 218 \, \alpha_i \alpha_j \alpha_k. \nonumber
\end{eqnarray*}
\indent
Let's start by rewriting (\ref{cOriginal}) as
\begin{eqnarray*}
c(t_1, \ldots, t_n) 	&=& 		\sum_{m=0}^n s(m) A_m,
\end{eqnarray*}
where $A_m$ is the sum over the product of all $m$-tuples of the $\alpha_i$:
\begin{eqnarray*}
A_0 &=& 1 \quad {\rm (by \,\, definition)}, \\
A_1 &=& \alpha_1 + \alpha_2 + \ldots + \alpha_n, \\
A_2 &=& \alpha_1 \alpha_2 + \ldots + \alpha_1 \alpha_n + \ldots \alpha_n \alpha_1+ \ldots + \alpha_n \alpha_{n-1}, \\
&\vdots& \\
A_n &=& \alpha_1 \ldots \alpha_n.
\end{eqnarray*}
\indent
We need to show that
\begin{equation}
\sum_{m=0}^n s(m) A_m		\leq 		s(n) \prod_{i=1}^n (\alpha_i+1).
\label{NN}
\end{equation}
To do so, consider the product 
\begin{equation*}
\prod_{i=1}^n (x + \alpha_i) = A_0 x^n + A_1 x^{n-1} + \ldots + A_n x^0.
\end{equation*}
Setting $x=1$, this gives
\begin{equation*}
\prod_{i=1}^n (\alpha_i + 1) = A_0 + A_1 + \ldots + A_n.
\end{equation*}

Substituting this into eq. (\ref{NN}) gives
\begin{equation*}
\sum_{m=0}^n s(m) A_m		\leq 		s(n) \sum_{m=0}^n A_m.
\end{equation*}
Since $s(m)$ is a non-decreasing function of $m$, $s(m) \leq s(n)$, and the above is true by inspection.
Therefore
\begin{equation*}
	c(t_1,\ldots,t_n) \leq s(n) \prod_{i=1}^n 2^{2^{t_i}}\!\!/2 = c_{\rm upper},
\end{equation*}
proving eq. (\ref{upperbound}).
\\ \\ \noindent {\sf\textbf{\textcolor{black}{Proof of lower bound on biological logics}}} \\
In addition to the upper bound, we can write a lower bound of a similar form.
To do so, we just take the last ($m = n$) term in eq. (\ref{cOriginal}) and substitute in $\alpha_i$.
This gives
\begin{equation*}
c_{\rm lower} = s(n) \prod_{i=1}^n \!\left(2^{2^{t_i}}\!\!/2 - 1\right) \leq c(t_1,\ldots,t_n),
\label{lower}
\end{equation*}
proving eq. (\ref{lowerbound}).
\\
\\ \begin{scriptsize}%
\noindent{\sf\textbf{Acknowledgements}} \\
{\sf\textbf{Funding:}}				{\sf This research was supported by a grant from bit.bio.}
{\sf\textbf{Competing interests:}}	{\sf The author declares that he has no competing interests.}
\end{scriptsize}

\end{small}

\vspace*{-0.4in}
\begin{thebibliography}{1}
\begin{scriptsize}
\bibitem{Gomez2008} 		C. Gomez et al., 							Control of segment number in vertebrate embryos,	 							Nature 			\textbf{454}, 	335 			(2008). 
\bibitem{Pawlowski2017} 		M. Pawlowski et al.,							Inducible and deterministic forward programming of human pluripotent stem cells 						
																into neurons, skeletal myocytes, and oligodendrocytes,							Stem Cell Reports	\textbf{8}, 		803 			(2017).
\bibitem{Kauffman1969} 		S.  A.  Kauffman,  							Metabolic  stability  and  epigenesis  in  randomly constructed genetic nets, 			J Theor Biol		22, 437 		(1969).
\bibitem{Huang2005}			S. Huang, G. Eichler, Y. Bar-Yam, and D. E. Ingber, 	Cell fates as high-dimensional attractor states of a complex gene regulatory network, 	Phys Rev Lett	 	\textbf{94}, 	128701 		(2005).
\bibitem{Socolar2003} 		J. E. Socolar and S. A. Kauffman, 				Scaling in ordered and critical random Boolean networks,  						Phys Rev Lett		\textbf{90},  	068702		(2003).
\bibitem{Samuelsson2003} 	B. Samuelsson and C. Troein, 					Superpolynomial growth in the number of attractors in Kauffman networks, 			Phys Rev Lett		\textbf{90}, 	098701 		(2003).
\bibitem{Shmulevich2004} 	I.  Shmulevich  and  S.  A. Kauffman,  			Activities and sensitivities in Boolean network models, 							Phys Rev Lett		\textbf{93}, 	048701		(2004).
\bibitem{Milo2002}			R. Milo et al., 								Network motifs: Simple building blocks of complex networks,						Science 			\textbf{298}, 	824 			(2002).
\bibitem{Milo2004}			R. Milo et al., 								Superfamilies of evolved and designed networks,								Science 			\textbf{303}, 	1538 		(2004).
\bibitem{Mangan2003} 		S. Mangan and U. Alon,						Structure and function of the feed-forward loop network motif,						P Natl Acad Sci USA \textbf{100}, 	980			(2003).
\bibitem{Ahnert2016} 	  	S. E. Ahnert and T. M. A. Fink, 					Form and function in gene regulatory networks 									J. R. Soc. Interface	\textbf{13}, 	20160179		(2016).
\bibitem{Whipple2021}		T. Whipple, 								''23 Mathematical Challenges'',	 											\emph{The Times},	June 2021.
\bibitem{Wagner2007} 		G. P. Wagner, M. Pavlicev, J. M. Cheverud,		The road to modularity,													Nat Rev Genet 		\textbf{8}, 		921 			(2007).
\bibitem{Decan2018}			A. Decan, T. Mens and E. Constantinou, 			On the evolution of technical lag in the npm package dependency network,"  			IEEE ICSME, 		p. 404					(2018).
\bibitem{Wolfram2002a}		S. Wolfram,								\emph{A New Kind of Science}												(Wolfram Media,					2002), p. 637.
\bibitem{Hannam2019}  		R. Hannam, R. Kuhn, and A. Annibale, 			Percolation in bipartite Boolean networks and its role in sustaining life, 				J Phys A 			\textbf{52}, 	334002		(2019).
\bibitem{Hannam2019b}  		R. Hannam, 								Cell states, fates and reprogramming, Ph.D. thesis, King's College London 													(2019).
\bibitem{Fink2022}			T. M. A. Fink and R. Hannam, 					Biological logics are restricted, 								submitted to 		Science Advances	 						(2021).
\bibitem{Wolfram2002b}		S. Wolfram,								\emph{A New Kind of Science}												(Wolfram Media,					2002), pp. 729, 1128.
\bibitem{Sloane}			N.  J.  A. Sloane, editor, 						The On-Line Encyclopedia of Integer Sequences, 								published electronically at https://oeis.org, 		2021.
\bibitem{Wagner2008}		A. Wagner,		 						Robustness and evolvability: a paradox resolved,								P R Soc B, 		\textbf{275}	91			(2008).
\bibitem{Buccitelli2020}   		C. Buccitelli and M. Selbach, 					mRNAs, proteins and the emerging principles of gene expression control, 				Nat. Rev. Genet	\textbf{21},	630 			(2020).
\bibitem{Newman2001} 		M. E. J. Newman, S. H. Strogatz, and D. J. Watts, 	Random graphs with arbitrary degree distributions and their applications, 				Phys Rev E		\textbf{64}, 	026118 		(2001).
\bibitem{Payne2015} 		J. L. Payne and A. Wagner, 					Mechanisms of mutational robustness in transcriptional regulation, 					Front Genet 		\textbf{6}, 		322 			(2015).
\bibitem{Dingle2018} 		K.  Dingle,  C.  Q.  Camargo,  and  A.  A.  Louis,  	Input-output maps are strongly biased towards simple outputs, 						Nat. Commun.		\textbf{9} 					(2018).
\bibitem{Johnston2022} 		I. G. Johnston et al., 							Symmetry and simplicity spontaneously emerge from the algorithmic nature of evolution,	P Natl Acad Sci USA	\textbf{119},	e2113883119 	(2022).
\bibitem{Raman2011} 	 	K. Raman and A. Wagner, 					The evolvability of programmable hardware, 									J. R. Soc. Interface	\textbf{8}, 		269 			(2011).
\bibitem{Bilke2001} 			S. Bilke and F. Sjunnesson, 					Stability of the Kauffman model,											Phys Rev E		\textbf{65}, 	016129 		(2001).
\bibitem{Reed2004}			M. Reed, 									Why is mathematical biology so hard?, 										Not Am Math Soc	\textbf{51}, 	338			(2004).
\bibitem{Reed2015} 			M. Reed, 									Mathematical biology is good for mathematics, 									Not Am Math Soc 	\textbf{62}, 	1172 		(2015).
%
\end{scriptsize}
\end{thebibliography}
\end{document}